\documentclass[twocolumn, aps,prl,reprint]{revtex4}
\usepackage[colorinlistoftodos]{todonotes}
\usepackage{amsfonts}
\usepackage{amssymb}
\usepackage{amsmath}
\usepackage[bookmarks = false, colorlinks = false, allcolors=blue]{hyperref} 
\usepackage{ulem}
\usepackage{marginnote}
\usepackage{tipa}
\usepackage{wasysym}
\usepackage{graphics,graphicx}
\usepackage{setspace}
\usepackage{tikz-cd}
\usepackage{tcolorbox}

\newcommand{\HH}{\mathcal{H}}

\newcommand{\Tr}{\mbox{Tr}}

\newcommand{\OO}{\mathcal{O}}

\newcommand\be{\begin{equation}}
\newcommand\ba{\begin{eqnarray}}
\newcommand\ee{\end{equation}}
\newcommand\ea{\end{eqnarray}}

\definecolor{lurple}{rgb}{0.8,0.0,0.6}
\definecolor{purple}{rgb}{0.5,0.0,0.3}
\definecolor{huh}{rgb}{0.0,0.6,0.8}
\definecolor{lorange}{rgb}{1,0.5,0}
\definecolor{orange}{rgb}{0.8,0.4,0}
\definecolor{dorange}{rgb}{0.6,0.2,0}
\definecolor{light-gray}{gray}{0.3}

\begin{document}
\title{A new look at the entanglement entropy of a single interval in a 2d CFT}
\author{Jennifer Lin}
\email{jylin04@gmail.com}
\affiliation{Future of Humanity Institute, University of Oxford, Oxford, OX2 0DJ, UK}

\begin{abstract}
In this note, I revisit the problem of computing the entanglement entropy of a single interval in the ground state of a 2d CFT. I write the leading-order result in three different ways: once by doing the replica trick with the $n$-replicated cylinder partition function computed in the ``closed string channel"; once by computing the $n$-replicated cylinder partition function in the ``open string channel", where the entanglement entropy can be related to the density of states in a boundary CFT defined on the interval; and for holographic CFTs, once as the log of a Plancherel measure for a certain noncompact quantum group.
 I comment on the implications for what the Ryu-Takayanagi area term might be counting in holographic CFTs. \end{abstract}

\maketitle

It is well-known that the entanglement entropy of a single interval in the ground state of a 2d CFT is given by
\be\label{e1}
S \sim \frac c 3 \log \frac L \epsilon 
\ee
at leading order in $\epsilon$, where $L$ is the length of the interval, $c$ is the central charge of the CFT and $\epsilon$ is a UV cutoff scale. The standard way to derive this starts from the geometric replica trick \cite{Holzhey:1994we}, \cite{Calabrese:2004eu}, according to which, for a state of a QFT that can be set up by cutting a Euclidean path integral on a manifold $\mathcal{M}$ at a moment of time-reflection symmetry, we can find the entanglement entropy across a region $A$ by first computing the partition function $Z_n$ of the QFT on an $n$-fold branched cover of $\mathcal{M}$ where we cyclically join successive copies along the region $A$. Then 
\be\label{e2}
S = -\lim_{n \rightarrow 1} \frac\partial{\partial n} \frac{Z_n}{Z_1^n}\,.
\ee
When $A$ is an interval in a 2d CFT, we can use conformal symmetry to further map the calculation of each such $Z_n$ to the calculation of a two-point function of twist operators with scaling dimension $\Delta = \frac c {24}(n - \frac 1 n)$. Since two-point functions in CFTs are totally fixed by conformal symmetry, this allows us to quickly derive \eqref{e1}.

By leaning as heavily on symmetry as it does though, this calculation, efficient though it is, obscures the question of what exactly is being entangled with what at the level of microscopic degrees of freedom within the CFT. This question has acquired new urgency in light of the ongoing project to understand the emergence of spacetime by studying entanglement entropy in holographic CFT's (see \cite{Rangamani:2016dms} for a review). In particular, \eqref{e1} is dual to the area of a Ryu-Takayanagi (RT) surface in AdS$_3$/CFT$_2$, and we would like to understand what the RT area term is counting.

In this note, I will revisit the problem of finding the entanglement entropy across an interval in a 2d CFT in three different ways. I will first review the usual story of how a careful use of the replica trick leads to $Z_n$ being a cylinder partition function that we can compute in the closed string channel. I will then show that if instead we compute $Z_n$ in the open string channel, we find that the entanglement entropy across an interval is dominated by modes at a particular ($L/\epsilon$-dependent) energy in the BCFTs that we implicitly split our CFT into when we do the entanglement calculation, assuming the exponential growth of states in the naive Cardy formula. Finally, I will show that for a CFT with a parametrically large central charge, we can rewrite \eqref{e1} as the log of a Plancherel measure for the quantum group $U_q(sl(2,R))$, evaluated on a particular ($L/\epsilon$-dependent) representation. This will provide evidence for a certain interpretation of what the Ryu-Takayanagi formula might mean in light of related recent work in SYK/JT holography \cite{Lin:2018xkj}, \cite{Lin:2021tb}.  

\section{Derivation of (1) in the ``closed string channel"}

Let us begin by more carefully setting up the replica trick calculation for the situation at hand. This section will closely follow \cite{Ohmori:2014eia} (though see also \cite{Cardy:2016fqc} for a  related discussion).

In \cite{Ohmori:2014eia}, the authors pointed out that entanglement entropy is only really ever defined w.r.t. a particular tensor product factorization $\HH = \HH_A \otimes \HH_{\bar A}$  of a theory's Hilbert space. To assign an entanglement entropy $S$ to the tensor factor $\HH_A$ in a state defined by the density matrix $\rho = |\psi\rangle\langle\psi|$, one can then trace over its complement $\HH_{\bar{A}}$ to get the reduced density matrix $\rho_A$, and take $S$ to be its von Neumann entropy, $S = -\Tr\, \rho_A \log \rho_A$\,. In a generic QFT however, there is no ``clear-cut factorization" of the Hilbert space by spatial subregion. So if we want to assign an entanglement entropy to a region $A$ in a QFT, we first have to explicitly force the QFT to factorize by cutting out a region of size $\epsilon$ around the boundary of $A$ and putting boundary conditions $a$ there. This operation can be represented by a linear map
\be\label{e3}
i: \HH \rightarrow \HH_{A,a} \otimes \HH_{\bar A, a}\,.
\ee
We can then compute {\it an} entanglement entropy {\it w.r.t. the particular choice of factorization map in \eqref{e3}.} Although this choice is often glossed over in the literature, \cite{Ohmori:2014eia} showed that it can leave a physical imprint on the universal parts of the entanglement entropy.

When applied to a single interval in a 2d CFT, the entangling region has just two endpoints, so the cutting map will take the form 
 \be\label{e4}
 i: \HH \rightarrow \HH_{a_1, A, a_2} \otimes \HH_{a_2, \bar A, a_1}\,.
 \ee
 To simplify the presentation below, I will assume that $a_1$ and $a_2$ are conformal boundary conditions. 

Now let us specialize to the ground state of the CFT, which we can set up by doing a Euclidean path integral on the half-plane $\tau < 0$. When we do the geometric replica trick for an interval of length $L$ in this case, the effect of the cutting map \eqref{e4} will be to excise a disk of size $\epsilon$ around each of the conical singularities on the replica manifold on which we compute $Z_n$, so that it ends up having the topology of a cylinder with boundary conditions $a_1$ and $a_2$ (instead of the sometimes-implicitly-assumed topology of a sphere). By applying a conformal transformation, we can further take $Z_n$ to live on a cylinder of circumference $2\pi$ and width $\ell/n$, where $\ell = \log (L/\epsilon)^2 + \OO((L/\epsilon)^{-1})$. 

This cylinder partition function can be computed in the ``closed string channel" as a propagator between the boundary states: 
 \be
 Z_n = \langle a_1 |\exp \left(\frac \ell n(\frac{c+\bar c}{24} - L_0 - \bar{L}_0) \right) | a_2\rangle\,.
 \ee
 If we insert a complete set of states $|k\rangle$ that can couple to $a_1$ and $a_2$, then the lowest-energy among them will dominate as $\epsilon \rightarrow 0$. 
 For a theory whose lowest-energy state has conformal dimension $\Delta_0 = 0$, this yields 
 \be
 Z_n \sim \langle a_1|0\rangle \langle 0|a_2\rangle \exp\left(\frac{c \ell}{12n} \right)\,,
 \ee
 from which plugging into \eqref{e2} gives the expected answer, \eqref{e1}. (These steps would also generalize to CFT's that don't have $\Delta_0 = 0$ if we replace $c \rightarrow c_{\rm eff} = c - 12\Delta_0$.)

Amusingly, note that if we define 
\be
p_k = Z_1^{-1} \langle a_1|k\rangle \exp\left(\ell (\frac c{12} - \Delta_k) \right) \langle k|a_2\rangle
\ee
to be the probability that the boundaries ``emit the mode $|k\rangle$", then by parametrizing $Z_n$ as
\be
Z_n = \sum_k (Z_1p_k)^{\frac 1 n}(\langle a_1|k\rangle \langle k|a_2\rangle)^{\frac{n-1}{n}}\,,
\ee
we can write the entanglement entropy as 
\be\label{e9}
S = \sum_k p_k(-\log p_k + \log \langle a_1|k\rangle + \log\langle k|a_2\rangle + 2\ell (\frac c {12} - \Delta_k))\,,
\ee
peeling off a Shannon term for the distribution over the $k$'s as well as analogs of the Affleck-Ludwig boundary entropy \cite{Affleck:1991tk} for the coupling of each mode $|k\rangle$ to the boundaries. 
(To recover \eqref{e1} from \eqref{e9}, note that $\left.p_k\right|_{|k\rangle = |0\rangle}\sim 1$ and that the last term in \eqref{e9} dominates as $\epsilon \rightarrow 0$.)
However, it is not clear that \eqref{e9} is physically meaningful since at this point $a_1$ and $a_2$ are just arbitrary boundary conditions that we put in to make the entanglement
 problem well-defined. 
This strategy of writing the entropy in terms of a suitable probability distribution, 
first introduced in \cite{Donnelly:2014gva}, will be more clearly meaningful in the parametrization that we turn to next.

\section{Derivation of (1) in the ``Open String Channel"} 

I will now redo the calculation but with $Z_n$ computed in the ``open string channel," where we take it to be a thermal partition function for the boundary CFT that lives on the interval of length $L$ with boundary conditions $a_1$ and $a_2$ at the ends of the interval. In this case, we find that if the boundary CFT has an exponential growth of states, then the modes responsible for the entanglement entropy across the interval will be peaked around a certain (length-dependent) energy. The idea of computing the replica partition function in the open string channel was suggested long ago by \cite{Susskind:1994sm}. 

In the open string channel, the natural way to write the $n$-replicated partition function is as
\begin{eqnarray}
Z_n &=& \Tr \exp \left(-\frac{2\pi^2 n}{\ell}\left(L_0 - \frac c {24}\right)\right) \\
&=& \int_\Delta d\Delta\,  \rho(\Delta) e^{-\frac{2\pi^2 n}{\ell}(\Delta - \frac{c}{24})} \label{e11}
\end{eqnarray}
where $\rho(\Delta)$ is the density of states of the boundary CFT. (See \cite{DiFrancesco:1997nk} for the conventions that we are using.)  
This density of states will depend in general on the microscopic details of the CFT, but at large $\Delta$ we can assume that it satisfies the naive Cardy formula \cite{Cardy:1986ie} (see \cite{Alba:2017bgn} for a recent derivation of the Cardy formula on the annulus), 
\be\label{e12}
\left.\rho(\Delta)\right |_{\Delta \rightarrow \infty}  \sim \exp\left(2\pi\sqrt{\frac c 6 (\Delta- \frac c{24})}\right)\,.
\ee
From here on I will ignore the shift by $c/24$ from the Casimir energy.

We now do the same trick as before, interpreting the summand of \eqref{e11} as a probability distribution. Defining 
\be\label{e13}
p_\Delta = Z_1^{-1}\rho(\Delta)e^{-\frac{2\pi^2}{\ell}(\Delta - \frac{c}{24})}\,,
\ee
we can write 
\be\label{e14}
Z_n = \int_\Delta d\Delta\, \rho(\Delta)^{1-n} (Z_1p_\Delta)^n\,,
\ee
from which the entanglement entropy takes the form
\be\label{e15}
S = \int_\Delta d\Delta\,  p_\Delta (-\log p_\Delta + \log \rho(\Delta))\,.
\ee

Eq. \eqref{e15} is true for any CFT. However, if we specialize to CFT's that satisfy \eqref{e12}, then the competition between the exponentials in \eqref{e12} and \eqref{e13} as $\epsilon \rightarrow 0$ means that $p_\Delta$ will be peaked at
\be\label{e16}
\Delta_{*} = \frac{c\ell^2}{24\pi^2}\,.
\ee

Evaluating the log term in \eqref{e15} on $\Delta_*$, we recover \eqref{e1} on the nose:
\be\label{e17}
S \sim \log \exp\left(2\pi \sqrt{\frac{c \Delta_*}{6}}\right) = \frac{c\ell}{6} \sim \frac c 3 \log \frac L\varepsilon\,.
\ee
For generic CFTs, this may just be a coincidence, since $S$ can get contributions from modes in a finite neighborhood of $\Delta_*$. However, if we take $c \gg 1$, then $p_\Delta$ will be  sharply peaked at $\Delta_*$ and the first approximation in \eqref{e17} will be a good approximation.

To summarize, by computing \eqref{e1} in the ``open string channel", we see that it primarily records the entanglement of modes centered around the level \eqref{e16} for any CFT that satisfies the Cardy formula. Moreover, for holographic CFTs with $c \gg 1$,  the entanglement entropy across an interval is well-approximated by the log of the density of states at the particular level \eqref{e16} in the BCFT.



\section{Continuum approximation of (15) for holographic CFTs}

For holographic CFTs, we can go one step further and use this relationship between the entanglement entropy and the density of states to rewrite the entanglement entropy across an interval as the log of a Plancherel measure for a continuum representation of the quantum group $U_q(sl(2,R))$. This will allow us to derive a 3d/2d analog of a result from SYK/JT holography \cite{Lin:2018xkj} that will suggest a certain interpretation for the Ryu-Takayanagi area term in AdS$_3$/CFT$_2$. I will derive the result in this section and then interpret it in the next. 

Namely, \cite{Jackson:2014nla} 
observed that the Cardy formula \eqref{e12} agrees with the $U_q(sl(2,R))$ Plancherel measure  
\be\label{e18}
\rho(p) = 4\sinh (2\pi bp)\sinh (2\pi b^{-1}p)
\ee
for $c \gg 1$, where $q = e^{i\pi b^2}$, and  the quantities in \eqref{e12} and \eqref{e18} are related via Liouville notation as 
\begin{eqnarray}
Q &=& b + b^{-1}, \\
c &=& 1 + 6 Q^2\,,	
\end{eqnarray}
and 
\be
\Delta = \frac 14 Q^2 + p^2\,. 
\ee

Redoing the ``open string" calculation with $\rho(p)$ in place of $\rho(\Delta)$ in \eqref{e11}, we find that 
\be\label{e22}
Z_n \sim \int dp\, \rho(p) e^{-\frac{2\pi^2 p^2n}{\ell}}
\ee
for large $c$.  In terms of 
\be\label{e23}
p_p = Z_1^{-1} \rho(p) e^{-\frac{2\pi^2p^2}{\ell}}\,,
\ee
we can write the entanglement entropy across an interval as
\begin{eqnarray}
S &\sim&  \int dp\, p_p(-\log p_p + \log \rho(p)) \label{e24} \\
&\sim& \log \rho(p)|_{p_*}\,, \label{e25}
\end{eqnarray}
where the new distribution \eqref{e23} is peaked at the value
\be\label{e26}
p_* = \sqrt{\Delta_*} =  \sqrt{\frac{c}{24}}\frac{\ell}{\pi}
\ee
of the Liouville momentum. Plugging \eqref{e26} into \eqref{e25}, we find again the universal answer 
\be\label{e27}
S \sim 2\pi(b + b^{-1})p_* = \frac {c\ell} 6 \sim \frac c 3 \log \frac L \epsilon\,.
\ee

\noindent Eq. \eqref{e24} resembles the main result from \cite{Lin:2018xkj}, where the entanglement entropy between the two SYK models in a Hartle-Hawking state of SYK/JT holography was  expressed in a similar functional form. 
	Here though, the result is approximate, with an error depending on the extent to which $\rho(p)$ is a good approximation for $\rho(\Delta)$.
	(See also \cite{Mcgough:2013gka} for earlier work in which the entropy of a BTZ black hole was written in terms of the $U_q(sl(2,R))$ Plancherel measure.) 


\section{Implications for Holography}

My motivation in revisiting the well-known entanglement entropy of a single interval in a 2d CFT was to try to better understand the meaning of the Ryu-Takayanagi (RT) formula in AdS$_3$/CFT$_2$. According to the RT formula \cite{Ryu:2006bv}, \cite{Ryu:2006ef}, in any state of a holographic CFT with an Einstein gravity dual, the entanglement entropy across a (possibly topologically nontrivial) 
region $A$ equals the area of the minimal surface $\gamma_A$ homologous to it in the bulk, 
to leading order in $1/G_N$: 
\be
S(A) = \frac{{\rm Area}(\gamma_A)}{4G_N}\,.
\ee
The RT formula has been verified in many settings (see \cite{Headrick:2019eth} for a review), proved in Euclidean signature \cite{Lewkowycz:2013nqa} and also generalized in many ways, most recently to a rule for computing the entanglement entropy of non-gravitational systems coupled to gravity that has led to significant progress on resolving the black hole information paradox (see \cite{Almheiri:2020cfm} for a review). However, we still don't understand why it is true from a canonical point of view.

One idea that was suggested some years ago is that the RT area term may be an analog of a $``\log \dim R$" edge term discovered while studying entanglement entropy in compact gauge theories, that quantifies constraints between a subsystem and its complement due to a (generalized) gauge symmetry \cite{Donnelly:2016auv}, \cite{Lin:2017uzr}. Our results \eqref{e24} - \eqref{e27} give some preliminary evidence that the ``group" in question might (at least approximately) be the quantum group $U_q(sl(2,R))$. Quantum groups are known to encode the duality data of associated CFTs, with irreps of the quantum group being mapped to conformal families of the associated CFT \cite{Moore:1989vd}, \cite{Gomez:1996az}. 
	The group $U_q(sl(2,R))$ in particular is supposed to capture the duality data for {\it all} possible Virasoro conformal blocks \cite{Ponsot:1999uf}. This suggests that each of the holographic BCFTs whose entanglement entropy we have been discussing should approximately contain Hilbert space sectors labeled by all possible conformal weights. 
		
In fact, the AdS$_3$/CFT$_2$ generalization of a more recent study of SYK/JT holography \cite{Lin:2021tb} suggests that each holographic BCFT should approximately have Hilbert space sectors labeled by {\it any number} of possible conformal weights, which we can parametrize with vectors of Liouville momenta.

A structure that would naturally lead to this conclusion is if the Hilbert space of each such BCFT 
could in some limit be identified with the space of Virasoro conformal blocks.
Then the conformal blocks might be the natural substrate to realize earlier pictures for what space itself might be made of such as tensor networks \cite{Swingle:2009bg}, bit threads \cite{Freedman:2016zud}, etc. 
(see also \cite{VanRaamsdonk:2018zws} for a suggestive picture in the limit that we chop the  CFT up into arbitrarily many BCFTs), with the RT area term being an edge term for gluing together legs of the conformal blocks in the different BCFTs. 
 In fact, \cite{Jackson:2014nla} had previously conjectured that the space of Virasoro conformal blocks should be a good approximation to the Hilbert space of any holographic CFT in the $\Delta > c/12$ regime.	

Going forward, we would like to get more evidence for this picture and to think about what we can use it for. For example, we would like to understand to what extent the picture 
is only approximately true. Later on, we would also like to understand how to get a smooth space{time} from this point of view, for which the edge modes of the CFT itself may be important (see \cite{Hung:2019bnq} for some work in this direction).

One way that we might hope to proceed is to keep on writing holographic entanglement results in novel ways, like in eqs. \eqref{e24} - \eqref{e27}. Unfortunately, this strategy 
seems hard to generalize already to the case of multiple intervals in the ground state of a 2d CFT, since we don't understand the modular flow with which to find an analog of the `open string channel' except in very special (free) cases \cite{Casini:2009vk}, \cite{Arias:2018tmw}. So to make further progress in this direction, we may have to come up with a different idea.

\bibliographystyle{ssg}
\bibliography{bhe}			
\end{document}